\documentclass[aps,twocolumn,showpacs,preprintnumbers,amsmath,amssymb]{revtex4}

\usepackage{epsf}
\newlength{\plotwidth}
\newcommand{\Gcr}{ {G_{\rm cr}} }
\newcommand{\Le}{ {\rm{Le}} }
\renewcommand{\Pr}{ {\rm{Pr}} }

\begin{document}


\title{Model Flames in the Boussinesq Limit: The Effects of Feedback}

\author{N. Vladimirova}
  \homepage{http://flash.uchicago.edu/~nata}
\author{R. Rosner}
\altaffiliation[also at]{ Departments of Astronomy \& Astrophysics and Physics,
The University of Chicago, Chicago, IL 60637}

\affiliation{ASCI Flash Center, Enrico Fermi Institute, 
The University of Chicago, Chicago, IL 60637}

\date{\today}

\begin{abstract}
  We have studied the fully nonlinear behavior of pre-mixed
  flames in a gravitationally stratified medium, subject to the
  Boussinesq approximation. Key results include the establishment of
  criterion for when such flames propagate as simple planar flames;
  elucidation of scaling laws for the effective flame speed; and a
  study of the stability properties of these flames. The simplicity of
  some of our scaling results suggests that analytical work may
  further advance our understandings of buoyant flames.
\end{abstract}


\maketitle

\section{Introduction}

In several areas of research, the feedback of a propagating diffusive
(pre-mixed combustion) flame on a fluid, and the consequent effects of
the flame itself, is of considerable interest. In the astrophysical
context, for example, the speedup of nuclear reaction fronts of this
type in the interior of white dwarf stars is thought to be one
possible way that such stars undergo thermonuclear disruption, e.g., a
Type Ia supernova (cf. 
\cite{Whelan73,Livne93,Khokhlov95,Niemeyer95,Reinecke99,Gamezo02}).
Much of the literature on this subject has focused
on the speedup of such flames for prescribed flows, and substantial
advances have been made in this regard recently \cite{Constantin00};
this is the ``kinematic'' problem, in which one seeks to establish
rigorous limits on flame speedup in the case in which there is no
feedback onto the flow. The aim of this paper is to study the simplest
case of feedback, namely that which occurs when a flame propagates
vertically, against the direction of gravity. As described
extensively in the previously cited literature, it is generally
believed that under such circumstances, the flame front is likely to
become distorted by the action of the Rayleigh-Taylor instability, and
thus achieves speedup; these calculations have been largely
illustrative, and based upon simulations using fully-compressible
fluid dynamics (e.g., \cite{Ropke02}) and fairly realistic nuclear
reaction networks.

Here, we focus on a much simpler problem;
we study such flames in the Boussinesq limit (leading to a far simpler
computational problem) and for highly simplified reaction terms
(avoiding the complexities of realistic nuclear reaction networks). In
this way, we are able to isolate the various effects which lead to
flame speedup, which is particularly important if one is to connect
such simulations to the extant analytical work on this subject (e.g.,
\cite{Audoly00,Berestycki02,Constantin00}). Indeed, an
important motivation for this work is to elucidate simple scaling laws
--- if they exist --- in order to suggest further analytical studies.

Our paper is structured as follows: In the next section, we describe
the specific physical problem we wish to study, establish the
equations to be solved, and describe the method of solution. In \S
III, we present our results, and in \S IV, we provide a summary and
discussion.


\section{The Problem}

The effect of gravity on the temperature distribution in a reacting
incompressible fluid with thermal diffusivity $\kappa$, viscosity
$\mu$, and density $\rho$ can be described by the set of
Navier-Stokes and advection-diffusion-reaction equations,
\begin{eqnarray}
    \rho \left[ \frac{\partial {\bf v} } {\partial t} +
    ({\bf v} \cdot \nabla) {\bf v}\right] &=&
    - \nabla p + \mu \nabla^2 {\bf v} + \rho {\bf g},
    \nonumber
    \\
    \frac{\partial T } {\partial t} + {\bf v} \cdot \nabla T &=&
    \kappa \nabla^2 T + R(T),
    \label{eqns_vT}
    \\
    \nabla \cdot {\bf v} &=& 0,
    \nonumber
\end{eqnarray}
where ${\bf v}$ is the fluid velocity and, without loss of generality,
the temperature $T$ has been normalized to satisfy $0 \leq T \leq 1$.
The thermal diffusivity and viscosity are assumed to be
temperature-independent, and density variations are assumed to be
small enough to be described by the Boussinesq model, e.g., $\rho(T) =
\rho_\circ + (\Delta \rho / \rho_\circ) T$. The model (\ref{eqns_vT})
can be derived from a more complete system under the assumption
of unity Lewis number ${\rm Le}$ (the ratio of thermal and material
diffusivities), and this article addresses only the ${\rm Le}=1$ 
case.

The Boussinesq model is the simplest system exibiting buoyancy effects
(and thus allowing for feedback to the flame) without introducing the
complexities associated with the presence of sound wave and
stratification of background ``atmosphere''. Because our intentions
is to elucidate basic priciples, rather than realistically modeling
specific physical situations, we view our approach as sufficient 
for the chosen task.

We consider a reaction term of Kolmogorov-Petrovskii-Piskunov
(KPP) type \cite{KPP}, of the form
\begin{equation}
R(T) = \alpha T(1-T)/4 \,,
\label{eq:KPP}
\end{equation}
where $\alpha$ is the (laminar) reaction rate. This reaction form has
an unstable fixed point at $T=0$, the ``unburned'' state, and a stable
one at $T=1$, the ``burned'' state. Thus a fluid element with positive
temperature will inevitably evolve to the burned state in a
characteristic time of order $1/\alpha$. As is well-known from the
combustion literature, the temperature equation from the system above
admits --- for a stationary fluid, and in the absence of gravity ---
one-dimensional solutions in the form of burning fronts propagating
with laminar burning speed $s_\circ$, and with characteristic flame
thickness $\delta$,
\begin{equation}
 s_\circ = \sqrt{\alpha \kappa} \, , \hspace{1cm}
 \delta = \sqrt{\alpha / \kappa } \, .
 \label{eq_laminar_flame_scaling}
\end{equation}
If it is further assumed that $T \rightarrow 1$ as $y \rightarrow -
\infty$, and $T \rightarrow 0$ as $y \rightarrow +\infty$, then the
front propagation is in the positive $y$ direction.

It is convenient to adopt the front thickness $\delta$ and the inverse
reaction rate $\alpha^{-1}$ as the units of distance and time
respectively. In these units the problem control parameters are the
Prandtl number $\Pr$ and the non-dimensional gravity $G$,
\begin{equation}
 \Pr = \frac{\nu}{\kappa} \,, \hspace{1cm}
  G = g \, \left( \frac{\Delta \rho}{\rho_\circ}\right) 
       \frac{\delta}{s_\circ^2} \,, 
\end{equation}
where $\nu$ is a kinematic viscosity $\nu=\mu/\rho_\circ$.
In addition, the system is characterized by a number of
length scales specifying the initial state, which are in
our case the dimensionless amplitude $A$ and the
dimensionless wavelength $L$ of the initial flame front
perturbation, $f(x) = a \cos ({2 \pi x} / {l})$,
\begin{equation}
  A=a/\delta\, , \hspace{1cm}
  L=l/\delta\, .
 \label{eq_defineAL}
\end{equation}

\begin{figure}
\begin{center}
\epsfxsize=\plotwidth
\mbox{\epsfbox{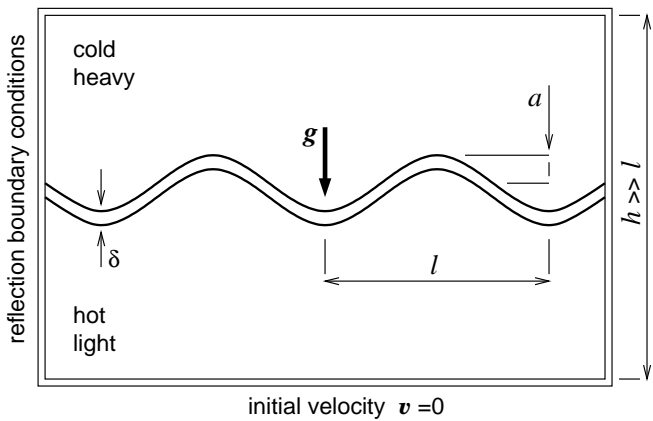}}
\end{center}
\caption{ A typical initial state of a flame calculation. }
\label{fig_intro}
\end{figure}

The vertical size of the computational domain was kept large so as to
avoid effects due to the upper and lower walls of the computational
box; in all cases, we have verified that such artifacts are
not present. For this reason, the box height does not enter as a
problem parameter. The initial velocities are set to zero, and most
computations were carried out for $\Pr=1$. A typical initial state of
our flame calculation is shown in Fig.~\ref{fig_intro}.

Because we focus on the two-dimensional problem, it is
convenient to re-write Eqns.~(\ref{eqns_vT}) in the stream
function and vorticity formulation in dimensionless form,
\begin{subequations}
\label{eq:omegaT}
\begin{equation}
  \frac{\partial \omega} {\partial t}  = \;
          - {\bf v} \cdot \nabla \omega \;
          + \;  \Pr \nabla^2 \omega \; - \;
           G \frac{\partial T} {\partial x} \,,
  \label{subeq:omega}
\end{equation}
\begin{equation}
  \frac{\partial T} {\partial t}
           = - {\bf v} \cdot \nabla T + 
               \nabla^2 T + \frac{1}{4}T(1-T) \, ,
  \label{subeq:T}
\end{equation}
\end{subequations}
using $\delta$ and $\delta/s_\circ$ as units of length and time
respectively. Here ${\bf v}$ is the non-dimensional velocity and $\omega$
is the non-dimensional vorticity ($\omega \equiv \nabla \times {\bf v}
= \nabla^2 \psi$). We solve the system (Eqns.~\ref{eq:omegaT})
numerically. The solution is advanced in time as follows: a third
order Adams-Bashforth integration in time advances $\omega$ and $T$,
where spatial derivatives of $\omega$ and $T$ are approximated by
fourth-order (explicit) finite differences. The subsequent elliptic
equation for $\psi$ is then solved by the bi-conjugate gradient method
with stabilization, using the {\tt AZTEC} library \cite{AZTEC}.
Finally we take derivatives of $\psi$ to update ${\bf v}$. 

The resolution of the simulations was chosen to fully resolve
the laminar flame structure. For the KPP reaction term (\ref{eq:KPP}),
the laminar flame thickness is approximately $12 \, \delta$, and the
grid spacing $\Delta x = \Delta y = 1$ (in the units of $\delta$) was
used in most of the computations. The laminar flame speed computed at
this resolution agrees with the theoretical value to within 1\%. This
corresponds to at least $16$~zones per wavelength (of the initial
perturbation) which is sufficient to resolve the flow. Most of the
computations were executed on a domain half the width of the initial
perturbation wavelength, and reflecting boundary conditions were applied.

Simulation times of $t=200$ -- $500$ (in units of $\delta/v_\circ$)
were required to measure the bulk burning rate, on computational grids
ranging from $8 \times 256$ for $L=16$ to $64 \times 2048$ for
$L=128$. Larger domains were necessary to obtain velocity fields
(e.g. $64 \times 3072$ for $L=128$), to avoid the influence of
boundary conditions at the top and bottom. Fortunately, only the
velocity in the reaction region affects the shape of the flame front
and, consequently, the bulk burning rate, so slight errors in
estimating velocities far away from the front due to upper and lower
boundaries do not affect our results. The comparison with linear
analysis was done using the same resolution and domain sizes up to
$512 \times 4096$.

By its nature, this study was comprised of a large number ($\sim 250$) of
simulations each representing a data point, as opposed to a close
examination of just a few simulations as in a case-study approach.
Confidence in the numerical accuracy was gained at the cost of a small
number of additional test simulations. Some simulations were repeated
using lower and higher resolutions, domains of different sizes in
vertical direction, and with several wavelengths across the width of the
domain. Special attention was devoted to simulations with different
Prandtl numbers, to ensure that both diffusive and viscous scales were
resolved.

Finally, a comment regarding the two-dimensional nature of our
simulations. In a recent study of the closely related Rayleigh-Taylor
instability \cite{Young01}, Young et al. specifically compared the
behavior of fingering and mixing in two- and three-dimensional flows,
with the result that while the specifics, e.g., finger growth rates,
were quite sensitive to dimensionality, the phenomenology nevertheless
turned out to be rather similar. Flames do however introduce a very
useful physical simplification into the Rayleigh-Taylor problem:
Because flames consume all density features at the flame front with
scales smaller than the flame thickness, the Rayleigh-Taylor problem
is ``regularized" by the burning process even in the limit of
vanishing viscosity. For this reason, a key difference between 2-D and
3-D -- namely, the difference between small-scale turbulent structures
in two and three dimensions -- is sharply reduced in the burning case.
The remaining difference between 2-D and 3-D is then mostly related to
the difference in propagation speed between buoyant parallel rolls
(the 2-D case) and buoyant tori (the 3-D case), with tori propagation
more quickly, i.e., we would expect 3-D flames to propagate more
quickly than 2-D flames, all other things being equal. We plan to
explore this point in future three-dimensional studies of flame
propagation.

\section{Results}

In this section, we discuss the results of our calculations, focusing
successively on the bulk burning rate, the evolution of the burning
travelling front, and the ultimate transition to a travelling (burning)
wave. Our central interest is in disentangling the dependence of the
flame behavior on the key control parameters of the problem.

\subsection{Travelling wave flame}

For a wide range of parameters, we were able to construct a
sufficiently large computational domain that we could observe
travelling waves of the temperature distribution, propagating with
constant speed. Depending on simulation parameters, the initial
perturbation either damps (e.g., the flame front flattens) or forms a
curved front. The flat front moves in the motion-free (in the
Boussinesq limit) fluid, has laminar front structure, and propagates
with the laminar front speed.

\begin{figure}
\begin{center}
\epsfxsize=\plotwidth
\mbox{\epsfbox{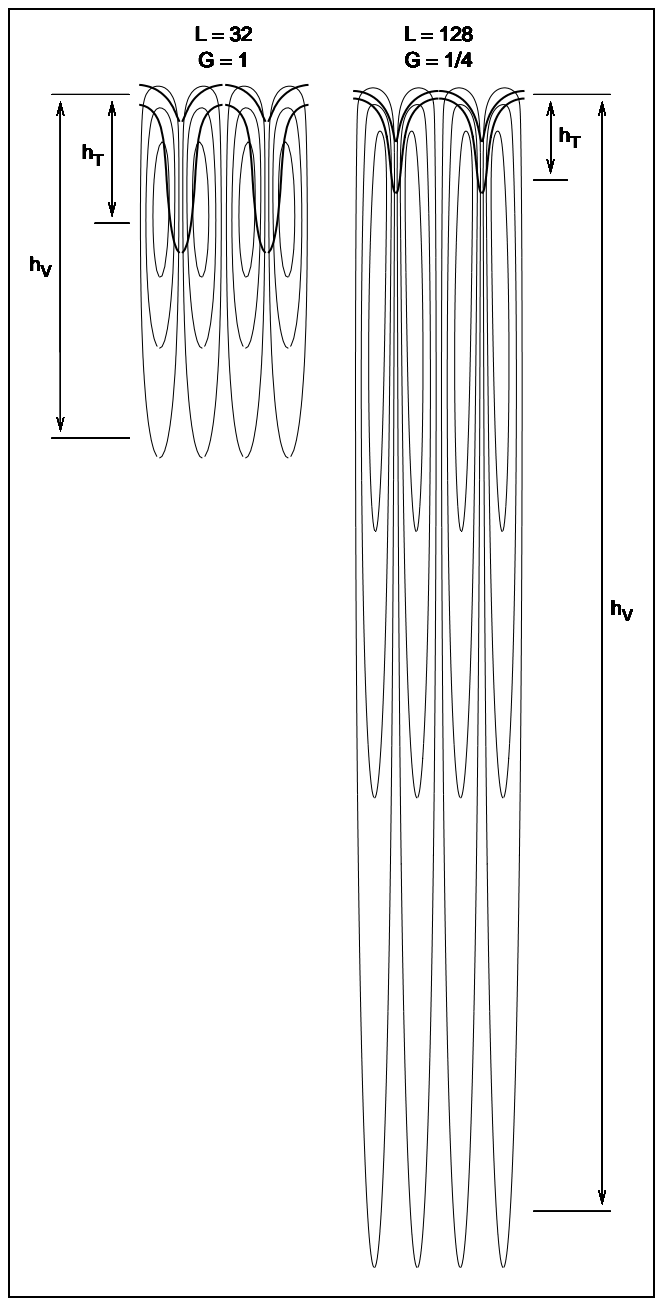}}
\end{center}
\caption{
  Travelling wave isotherms ($T=0.1$ and $T=0.9$) and streamlines for
  two system with different simulation parameters. Note that the
  system on the right has been rescaled by a factor of $1/4$ both
  horizontally and vertically.}
\label{fig_levels}
\end{figure}

The typical curved front is shown in Fig.~\ref{fig_levels}; it has the
wavelength of the initial perturbation and is characterized by narrow
dips (lower apexes), where the cold fluid falls into the hot region,
and by wide tips (upper apexes), where buoyant hot fluid rises into
the cold fluid. In the initial stages, the evolution pattern is similar
to bubble and spike formation during the Rayleigh-Taylor instability
\cite{Chandrasekhar61,Landau87}; in later stages, small scale
structures are consumed by the flame, and, finally, the flame evolves
toward the travelling wave solution shown in Fig.~\ref{fig_levels}.
The shape of the stable front is determined by gravity, $G$, and
wavelength, $L$, and can be characterized by two vertical length
scales associated with the spatial temperature variation ($h_T$) and
the spatial velocity variation ($h_V$) of the flame. The speed of the
curved front is always higher than the laminar flame speed, because of
the increase in the flame front area and transport. Finally, we
notice that the streamlines in Fig.~\ref{fig_levels} indicate that
the flow underlying the propagating flame is characterized by rolls
propagating upward.

One of our primary interests is to quantify the effects of variations
in wavelength and gravity on the flame speed. It is convenient to
define the speed of the travelling wave flame by the bulk burning rate
\cite{Constantin00},
\begin{equation}
  s(t) = \frac{1}{l} 
         \int^l_0 \frac{\partial T(x,y,t)}{\partial t} ~dx~dy \,;
\end{equation}
this definition has the considerable advantage that it reduces to the
standard definition of the flame speed when the flame is well-defined,
and it is accurate to measure even for cases where the burning front
itself is not well-defined. Henceforth we refer to it simply as the
flame speed.

\begin{figure}
\begin{center}
\epsfxsize=\plotwidth
\mbox{\epsfbox{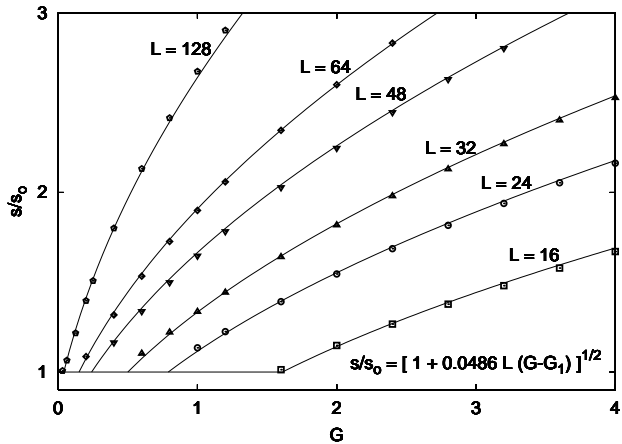}}
\end{center}
\caption{Bulk burning rate (travelling wave speed) $s$ as function of
wavelength $L$ for different values of gravity $G$.}
\label{fig_flame_speed}
\end{figure}

\begin{figure}
\begin{center}
\epsfxsize=\plotwidth
\mbox{\epsfbox{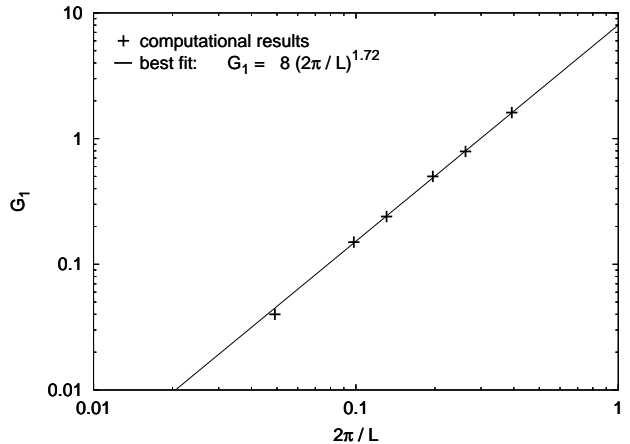}}
\end{center}
\caption{Transitional point $G_1$ as a function of wavelength.}
\label{fig_transitional_gravity}
\end{figure}

\begin{figure}[b]
\begin{center}
\epsfxsize=\plotwidth
\mbox{\epsfbox{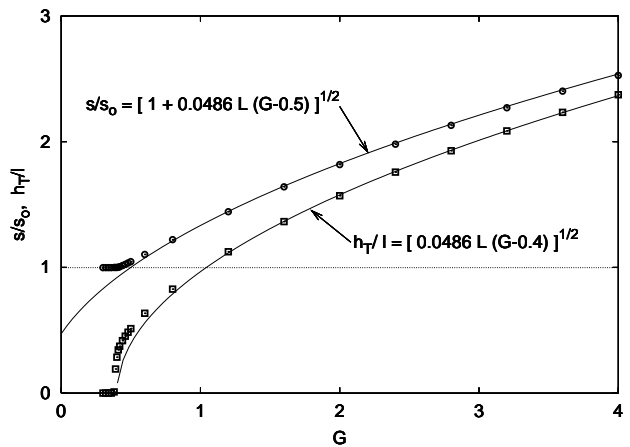}}
\end{center}
\caption{Amplitude of the stable front as function of gravity for
  the wavelength $L=32$. The scaling relations shown here are
  discussed in the text.}
\label{fig_ht}
\end{figure}

Our first result (shown in Fig.~\ref{fig_flame_speed}) is
that the flame speed increases with wavelength $L$ and with the
gravitational acceleration $G$, and is independent of the initial
perturbation amplitude $A$. More specifically, the flame becomes
planar and moves at the laminar speed ($s=s_\circ$) if $G$ is smaller
than some critical value $\Gcr$; if $G$ lies above this critical
value, the flame speed can be fit by the expression,
\begin{equation}
  s = s_\circ \sqrt{ 1 + k_1 (G-G_1) L } \,,
  \label{eq_flame_speed}
\end{equation}
where $k_1 \approx 0.0486$ is obtained from measurements derived from
the simulation data. The second tuning parameter, $G_1$, was found to
be a function of the perturbation wavelength
(Fig.~\ref{fig_transitional_gravity}), $G_1 = 8 (2\pi /L)^{1.72}$.
For a relatively wide range of parameters, Eq.~(\ref{eq_flame_speed})
describes experimental data well, but must be applied with caution
near the cusp at $G=G_1$ shown in Fig.~\ref{fig_flame_speed}.
Roughly speaking, this cusp can be interpreted as the transition
between the planar and curved flame regimes, $G_1 \approx \Gcr$;
closer investigation of the transition region shows that $\Gcr < G_1$,
and that the fit (Eq.~\ref{eq_flame_speed}) underestimates the flame
speed in this transition region (Fig.~\ref{fig_ht}).

The behavior near the transition is discussed in detail in the
theoretical work carried out by Berestycki, Kamin \& Sivashinsky
\cite{Berestycki02}. They derive the one-dimensional evolution
equation for the front interface $y(x)$ and prove mathematically the
following properties of $y(x)$, relevant to our case: (1)~the
existence of $\Gcr \sim (2\pi/L)^2 $ such that there is no nontrivial
solution for $G<\Gcr$ (i.e. the front is flat for $G<\Gcr$); (2)~the
existence of $\Gcr^* = 4\Gcr$ such that for $G > \Gcr^*$ there are two
symmetrical (curved) solutions $y^+(x)$ and $y^-(x)$ which are stable,
and any other solution including the trivial is unstable;
(3)~metastability of any solution except $y^+(x)$ and $y^-(x)$ in the
range $\Gcr < G < \Gcr^*$, and convergence of this metastable solution
to either $y^+(x)$ or $y^-(x)$. Also, based on the derivation in
\cite{Berestycki02}, it can be shown that the flame speed in the
metastable regime scales as follows \cite{Kiselev},
\begin{equation}
  (s/s_\circ - 1) \propto (G-\Gcr)^2 ~~
  {\rm as}~~ (G-\Gcr) \rightarrow 0 \,.
\end{equation}

Our simulations confirm the dramatic increase of stabilization times
close to the critical gravity value $\Gcr$. For this reason, it is
very difficult to obtain reliable results regarding the flame speed in
this transition regime. Even detecting the critical point takes
significant computational effort (Fig.~\ref{fig_ht}); measuring the
velocity, which in this parameter regime differs from $s_\circ$ by a
very small amount, is harder still. 

However, the transition is sharper and is easier to see when studying
the vertical distance between the upper and lower apexes of the flame,
$h_T$, measured by the expression,
\begin{equation}
  h_T = \int_{-\infty}^{\infty} \left( T(0)-T(l/2) \right) dy \,.
\end{equation}
In the limit of large wavelengths ($L \gg 1$), the transition occurs
at small values of gravity, and the flame speed is determined by a
single parameter, the product $LG$. If, in addition, the product $LG$
is large, the flame speed scales as $s/s_\circ \approx 0.22\sqrt{LG}$.
This result is in good agreement with the rising bubble model
\cite{Bychkov00} which, in the Boussinesq limit, predicts $s/s_\circ =
\sqrt{LG/6\pi} \approx 0.23 \sqrt{LG}$ for a 2-D open bubble
\cite{Layzer55}. We further observe that in the large wavelength limit,
the $h_T/l$ ratio obeys the same scaling (Fig.~\ref{fig_ht}).

\begin{figure}
\begin{center}
\epsfxsize=\plotwidth
\mbox{\epsfbox{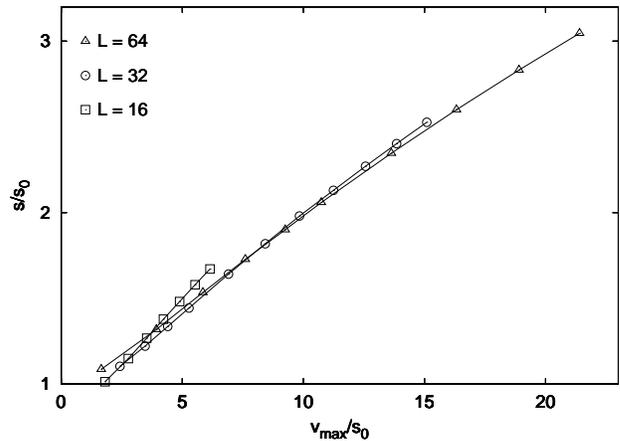}}
\end{center}
\caption{The flame speed as function of maximal flow velocity.}
\label{fig_vmax}
\end{figure}

We note that the flame structure shares features of flame propagation
from both shear and cellular flow. For instance, the temperature
distribution closely resembles that of a flame distorted by a shear
flow, while the velocity distribution resembles that inside an
infinitely tall cell. The flame speed in the shear and cellular flow
is determined by the flow speed and by the length scale of the flow
(period of shear or cell size) \cite{Vladimirova03}. In particular,
in both cases the flow speed scales with maximum flow velocity as $s
\propto v_{max}^n$, with $n=1$ for burning in the shear flow and
$n=1/4$ for burning in the cellular flow. Similarly, we have tried to
determine whether the flame speed relates to the maximum velocity of
the flow when flow and flame are coupled through the Boussinesq
model. The available data (shown in Fig.~\ref{fig_vmax}) do not
demonstrate a power law dependence with a single well-defined
power. Furthermore, the dependence on $L$ is not as dramatic as in the
cases of shear or cellular flows.

\subsection{The thin front limit}

The thin front limit is particularly important for developing models
of flame behavior. For many applications --- especially in
astrophysics --- resolving flames (by direct simulation) is
prohibitively expensive, and understanding flame propagation in the
limit in which the flame front becomes indefinitely thin (when
compared to other length scales of the application) is critical for
designing flame models. Of course, this same limit is of intrinsic
mathematical interest.

Particularly important is the dependence of the flame speed on the
wavelength of the front perturbation in the thin front limit. We have
already pointed out that instabilities with larger wavelengths have
higher travelling wave speeds, so that eventually the instability with
the largest wavelength allowed by the system dominates. (In our
non-dimensionalization, this is the instability with the highest ratio
of wavelength to laminar front width).

In this context, it is convenient to switch from our ``laminar flame
units'' to the so-called ``G-equation units''. The G-equation is a
model for reactive systems where very low thermal diffusivity is
exactly balanced by high reaction rate (see e.g., \cite{Pelce88}). The
diffusion and reaction terms in the temperature equation are replaced
by a term proportional to the temperature gradient,
\[
    \frac{\partial T } {\partial t} + {\bf v} \cdot \nabla T =
    s_\circ | \nabla T | \, ,
\]
so that the front propagates normal to itself at the laminar flame
speed $s_\circ$. The Boussinesq fluid model, combined with the
G-equation flame model, has the following physical parameters: (1)
flow length scale $l$, (2) laminar flame speed $s_\circ$, (3) gravity
$g$, and (4) fluid viscosity $\nu$. Choosing $l$ and $l/s_\circ$ as
the length and time units, the governing non-dimensional parameters
are $\tilde g = g l / s_\circ^2$ and $\tilde \nu = \nu/ls_\circ$; the
corresponding parameters in the laminar flame unit system are $\tilde
g = LG$ and $\tilde \nu =\Pr /L$. Note that in the limit $L
\rightarrow \infty$ while keeping $\Pr=1$, the Navier-Stokes equation
becomes the Euler equation and $\tilde \nu \rightarrow 0$, leaving
only one parameter in the system, $\tilde g = ( \Delta \rho /
\rho_\circ ) g l / s_\circ^2 = LG$.

\begin{table}[b]
\caption{\label{tab:table1}
Three simulations with $LG=32$ discussed in the text.
}
\begin{ruledtabular}
\begin{tabular}{crcrrc}
setup & $L$ & $G$ & $s/s_\circ$ & $h_T/l$ & $v_{\rm max}/s_\circ$ \\
\hline
(a) &  32 &  1   & 1.34 & 0.98 & 4.40 \\
(b) & 128 & 1/4  & 1.51 & 0.83 & 5.06 \\
(c) & 256 & 1/8  & 1.57 & 0.75 & 5.01 \\
\end{tabular}
\end{ruledtabular}
\end{table}

In our simulations $\Pr=1$, so it is not surprising that for large
$L$, almost all aspects of the system are well characterized by the
$LG$ product alone. For example, the formula for the bulk burning
rate, $s/s_\circ=\sqrt{1+k_1 LG}$ for $G>G_1$ well describes our
experimental results. Next, consider the travelling wave solutions
shown in Fig.~\ref{fig_levels}; these two systems have the same $LG$
product, with $L=32$ and $L=128$, and move at the speed
$s/s_\circ=1.34$ and $s/s_\circ=1.51$ respectively. The wavelength
here is comparable to the laminar front thickness (indicated by the
two limiting isotherms $T=0.1$ and $T=0.9$). Still, the front shape
as well as flame speed and fluid velocities are very similar.

\begin{figure}
\begin{center}
\epsfxsize=\plotwidth
\mbox{\epsfbox{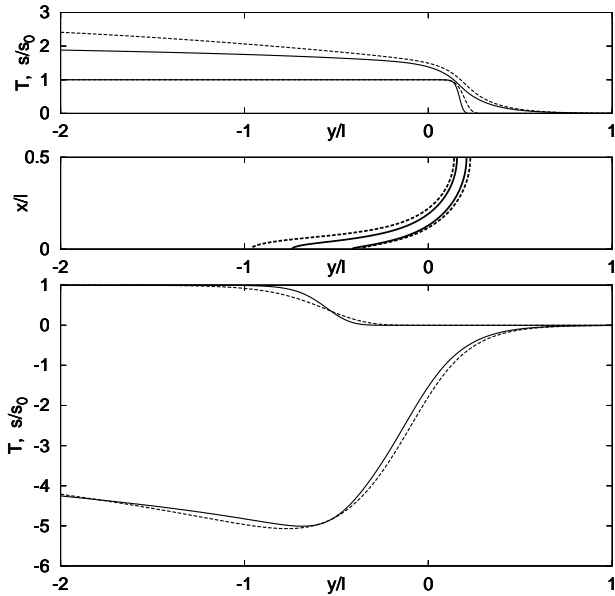}}
\end{center}
\caption{
  Travelling wave solution for two systems with $LG=32$, with
  $L=128$ (dashed lines), and with $L=256$ (solid lines).
  The isotherms $T=0.1$ and $T=0.9$ are shown in the middle panel.
  The top panel shows the temperature profiles and vertical velocities
  (along $\bf{g}$) at $x=0.5$ (upper flame front apex); the bottom
  panel plot shows the same things at $x=0$ (lower flame front apex).}
\label{fig_walls}
\end{figure}

One can see similarity more clearly in Fig.~\ref{fig_walls} (middle
panel), which compares systems with $L=128$ and $L=256$. The agreement
between bulk burning rates is very good ($s/s_\circ=1.51$ and
$s/s_\circ=1.57$). The match between the two integral measures $h_T$
is weaker ($h_T/l=0.83$ and $h_T/l=0.75$), suggesting that the systems
in consideration are still far from the infinitely thin front limit,
but this is apparent from the distance between limiting isotherms. We
have also compared the temperature and velocity profiles at the upper
and lower apexes of the flame (Fig.~\ref{fig_walls}, the top and the
bottom panels). The velocity is --- as expected --- essentially zero
well ahead of the temperature front, but significant motion extends
far behind it; the absolute maximum velocity is located in the
vicinity of the lower apex and is related to the bulk burning rate
(Fig.~\ref{fig_vmax}). By examing the detailed velocity profiles we
find that velocities at the flame front also obey the $LG$ product
scaling and, together with the temperature distribution, determine
the bulk burning rate. However, the velocities well behind the front
can be quite different for two systems with the same $LG$ product (cf.
Fig.~\ref{fig_levels}).

\begin{figure}
\begin{center}
\epsfxsize=\plotwidth
\mbox{\epsfbox{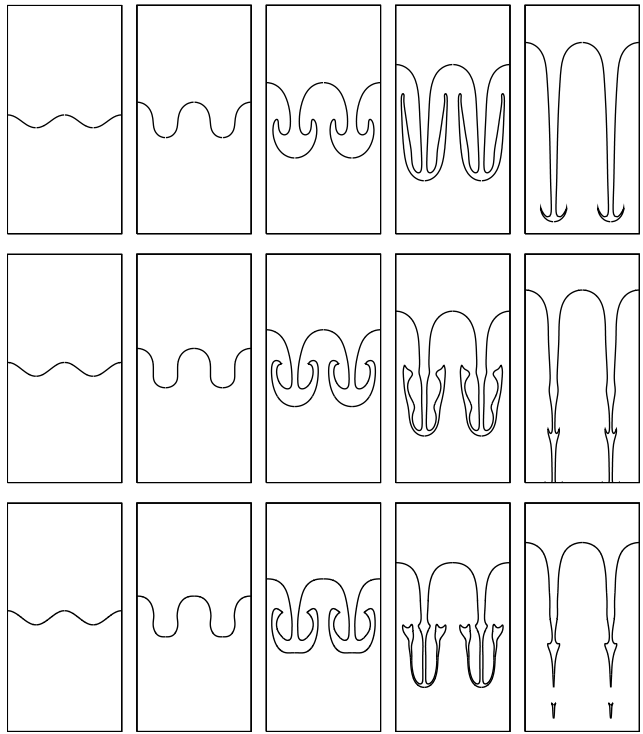}}
\end{center}
\caption{
  The isotherm $T=0.5$ during the instability growth phase, shown for
  three systems with $LG=512$ but different $L$ (top: $L=64$, middle:
  $L=128$, bottom: $L=256$). The initial amplitude is $a/l= 1/8$, and
  snapshots are taken at times $t (s_\circ/l) = 0, \, 1/16, \, 2/16,
  \, 3/16, \, 4/16 \, .$}
\label{fig_interface_similarity}
\end{figure}

Finally, consider the temperature during the instability growth phase,
shown for three different cases (with $LG=512$) in
Fig.~\ref{fig_interface_similarity}. Although the wavelength to
laminar front thickness ratio affects small scale features, we again
clearly see the similarity scaling connecting these solutions.

As we have shown above, the dependence on a single parameter, namely
the $LG$ product, in the infinitely thin front limit follows from
dimensional analysis; and for reasonably thin fronts, we were able to
confirm the $LG$ product scaling. At the same time, we noticed that
the length of the velocity variation, $h_V$, does not scale with $LG
\equiv \tilde{g}$. It is reasonable to assume that $h_V$ is controlled
by the other parameter, namely, the non-dimensional viscosity $\tilde
\nu = \Pr /L$, which is essentially zero in the thin flame limit. One
can understand this as follows.

\begin{figure}
\begin{center}
\epsfxsize=\plotwidth
\mbox{\epsfbox{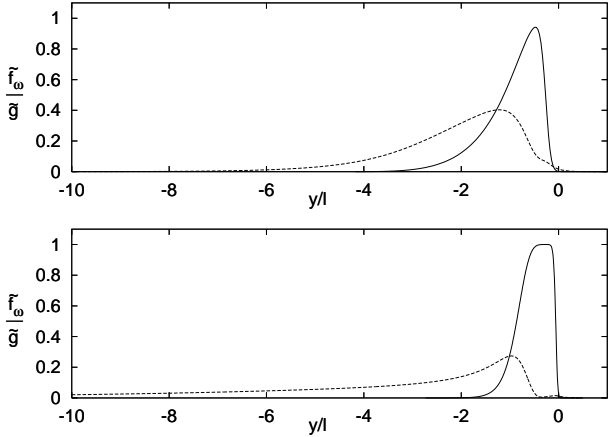}}
\end{center}
\caption{Vorticity generation in the roll (solid line)
  and vorticity fluxes through the separatrices between the rolls
  (dashed line), as a function of height $y$ for a flame with $L=32$
  and $G=1$ (top panel) and for a flame with $L=128$ and $G=1/4$
  (bottom panel). The areas below the solid and dashed lines are
  equal to $h_T/l$.}
\label{fig_vorticity_balance}
\end{figure}

From Eq.~(\ref{subeq:omega}), we can see that vorticity is generated
in the regions with significant temperature gradients, e.g., on the
scale $h_T$, and is advected by the flow on spatial scales of order
$h_V$. Thus, positive vorticity is generated in the domains
$nl<x<(n+1/2)l$, while negative vorticity is generated in the domains
$(n+1/2)l<x<(n+1)l$; however, the total (signed) vorticity in the
domain is conserved. Diffusion of vorticity occurs predominantly
across the boundaries $x=nl/2$. More directly, it is straightforward
to integrate the vorticity equation (Eq.~\ref{subeq:omega}) over the
area $nl<x<(n+1/2)l$ to obtain the vorticity balance,
\begin{eqnarray}
  \dot{\Omega} &=&  \tilde{g} s_\circ^2 \frac{h_T}{l} = \nonumber \\
               &=& \tilde{\nu} s_\circ^2
        \int_{-\infty}^{\infty} 
        \left[ \frac{\partial^2 \tilde{v}_y}{\partial \tilde{x}^2} 
        \right]_{\tilde{x}=0} +
        \left[ \frac{\partial^2 \tilde{v}_y}{\partial \tilde{x}^2} 
        \right]_{\tilde{x}=1/2}
        d\tilde{y} \,. \nonumber
\end{eqnarray}
Here $\dot{\Omega}$ is the total vorticity generated in the roll
$nl<x<(n+1/2)l$, and diffused through its boundaries. In
Fig.~\ref{fig_vorticity_balance} we have plotted the non-dimensional
vorticity generation, averaged in the area element $(l/2,\Delta y)$,
$\tilde{f}_\omega = \tilde{g}(l/s_\circ^2)(\Delta\dot{\Omega}/\Delta
y)$, and corresponding fluxes across the roll boundaries. Note that
only diffusion can lead to vorticity transport across roll boundaries
because the transverse flow vanishes identically at the separatrices.

In other words, in the thin flame limit, vorticity generation depends
on the $LG$ product, but not on the viscosity; however, in steady
state, we know that vorticity generation and destruction must balance
exactly.  Since the vorticity destruction depends on the diffusion
term $\Pr \nabla^2 \omega$, which decreases as $L$ increases, balance
can only be achieved if the length of the vorticity diffusion region
(i.e. the separatrix separating adjacent rolls) lengthens. Thus, we
expect $h_V$ to scale inversely with $\tilde{\nu} = \Pr /L$. Indeed,
we expect $h_V \rightarrow \infty$ as $\Pr \rightarrow 0$.


\subsection{Comparison with linear stability analyses}

A thorough analysis of the linear behavior of our system was
presented by Zeldovich et al. \cite{Zeldovich85}; in this subsection,
we compare our results with theirs.

The simplest case studied is the so-called Landau-Darrieus
instability, in which the flame is considered as a simple gas dynamic
discontinuity \cite{Darries38,Landau44}. The fluid on either side of
the discontinuity is assumed to obey the Euler equation; the fluid is
assumed to be incompressible; there is no temperature evolution
equation; and the front is assumed to move normal to itself with a
given laminar speed.  The important parameter is the degree of thermal
expansion, $\theta \equiv \rho_{\rm fuel} / \rho_{\rm ash}$, across
the flame front. The resulting instability growth rate is proportional
to the product of the laminar flame speed and the wavenumber of the
front perturbation, with a coefficient of proportionality depending on
$\theta$. For $\theta = 1$, which corresponds to the Boussinesq limit,
the growth rate is identically equal to zero.

The Landau-Darrieus model is however not valid for wavelengths short
compared to the flame thickness, for which it predicts the largest
growth rate; this deficiency was resolved by Markstein
\cite{Markstein64}, who introduced an empirical ``curvature
correction'' for the flame speed within the context of the
Landau-Darrieus model. One consequence of this correction is that the
instability is suppressed for wavelengths shorter than a specific
critical cutoff wavelength, while for wavelengths much larger than
this cutoff lengthscale the growth rate approaches zero as $1/L$, just
as in the Landau-Darrieus model.

\begin{figure}
\begin{center}
\epsfxsize=\plotwidth
\mbox{\epsfbox{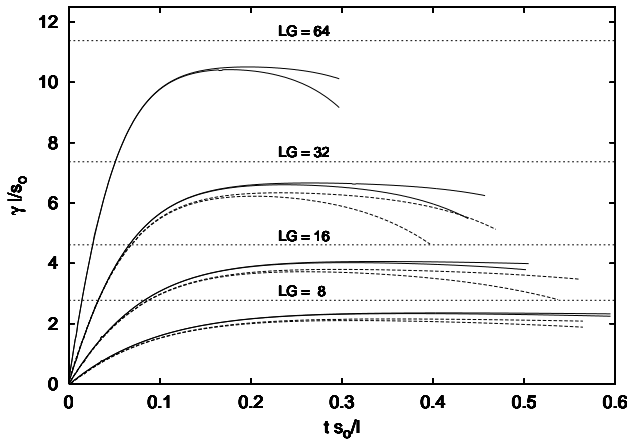}}
\end{center}
\caption{Growth exponent for a single mode and two initial amplitudes
  $A\equiv a/\delta=4$ and $A=8$. The dimensionless wavelengths are
  $L\equiv l/\delta=512$ (dashed) and $L=1024$ (solid). The dotted
  lines correspond to the linear stability analysis prediction,
  Eq.(\ref{eq_growth_thin_front}).}
\label{fig_expo_devel}
\end{figure}

Gravity can be introduced in this type of model in a very similar way,
as shown by Zeldovich et al. \cite{Zeldovich85}. Rewriting their result
in our notation, and taking into account $\theta = 1$ and $\Le=1$
(which leads to the Markstein curvature correction constant being set
equal to unity), we can reduce their final result to the following
expression for the growth rate,
\begin{equation}
 \gamma =  \frac{s_\circ}{2\delta} k
 \left[ \sqrt{ 1 + k(k - 2) + 2G/k } - 1 - k \, \right],
\label{eq_markstein}
\end{equation}
where $k=2\pi/L$. A more elaborate model for the flame, introduced by
Pelc\'e \& Clavin \cite{Pelce82}, avoids the empirical curvature
correction constant and, in the Boussinesq limit, gives the growth
rate expression
\begin{equation}
 \gamma =  \frac{s_\circ}{2\delta} k
 \left[ \sqrt{ 1 + 2G/k } - 1 - k - \frac{k}{\sqrt{1 + 2G/k}} \right] \, .
\label{eq_pelce}
\end{equation}
In the limit of thin fronts, $L \gg 1$, both models reduce to the same
expression, which also recovers the $LG$ similarity scaling already
discussed above,
\begin{equation}
\gamma \frac{l}{s_\circ} = 
 \pi \left( \sqrt{ 1 + \frac{1}{\pi} LG } - 1 \right).
\label{eq_growth_thin_front}
\end{equation}

\begin{figure}
\begin{center}
\epsfxsize=\plotwidth
\mbox{\epsfbox{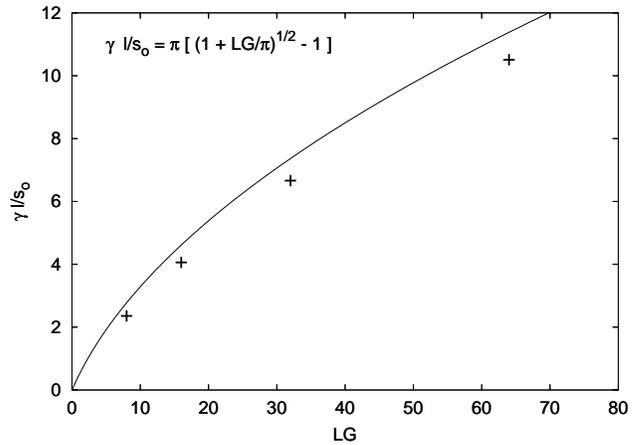}}
\end{center}
\caption{Growth exponent for a single mode
  measured at the maximum for $A=4$ and $L=1024$. The solid line
  corresponds to the linear stability analysis prediction,
  Eq.(\ref{eq_growth_thin_front}).}
\label{fig_expo}
\end{figure}

To compare our calculations with this result, we have computed the
growth rate for a single wavelength for a system with $L = 512$ and
$L= 1024$ (see Fig.~\ref{fig_expo_devel}). The growth rates predicted
by Eq.(\ref{eq_growth_thin_front}) are shown as horizontal dotted
lines for each $LG$ product. An ideal system in the linear regime
would have a constant growth rate; in our simulations we observe an
essentially time-independent growth rate only after some transitional
period, $t < 0.1 \, l/s_\circ$, and before the flame stabilization
time, which depends on parameters. The transitional period at the
beginning of our simulations can be explained by artificial initial
conditions, e.g. zero velocity and prescribed temperature profile
across interface. The decrease in the growth rate at later times is
related to the stabilization of the flame front. Naturally, the
faster-growing instabilities with higher $LG$ product and the systems
starting with larger initial amplitudes reach the steady-state more
quickly. In addition, we observe the influence of the finite flame
thickness --- plots with $L=1024$ approache closer to the infinitely
thin limit than plots with $L=512$. But in spite of the finite flame
thickness and non-zero viscosity, one can clearly see the similarity
scaling on $LG$ and good agreement with theory (Fig.~\ref{fig_expo}).

\begin{figure}
\begin{center}
\epsfxsize=\plotwidth
\mbox{\epsfbox{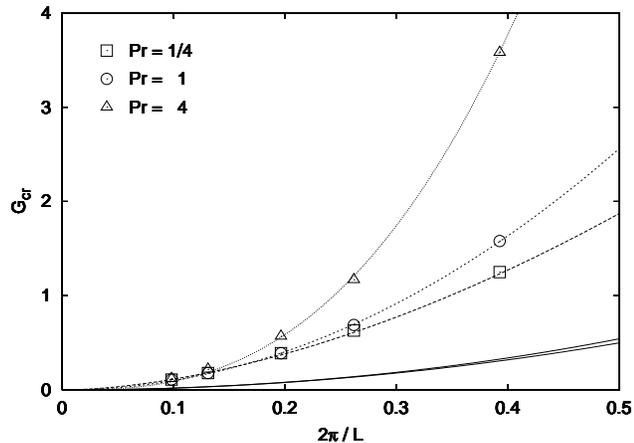}}
\end{center}
\caption{
  Critical gravity $\Gcr$ for different values of $\Pr=\nu/\kappa$.
  Our results are fit with power laws, in the form $\Gcr=C(2\pi/L)^n$,
  with measured values of $n=2.68$, $C=43.94$ for $\Pr=4$; $n=2.01$,
  $C=10.31$ for $\Pr=1$; and $n=1.73$, $C=6.21$ for $\Pr=1/4$. The two
  solid lines are provided by inviscid theory ($\Pr=0$), corresponding
  to the Markstein and the Pelc\'e \& Clavin models.}
\label{fig_prandtl}
\end{figure}

In order to obtain the stability condition, we set $\gamma = 0$ in the
expressions (\ref{eq_markstein}) and (\ref{eq_pelce}), and obtain
\begin{equation}
  G_{\rm cr} = 2 k^2 \,
\end{equation}
for the Markstein model, and
\begin{equation}
  G_{\rm cr} = \frac{k}{2} \left[ \frac{1}{4} 
   \left( 1+k+\sqrt{(1+k)^2 +4k} \right)^2 - 1\right] 
\end{equation}
for the Pelc\'e and Clavin model. We emphasize that both of these
models assume an inviscid fluid, while viscosity is present in our
simulations. In Fig.~\ref{fig_prandtl} critical gravities derived
using both models are plotted next to numerical simulations data for
different Prandtl numbers.

Similarly, we can consider the relation between the instability growth
rate and the amplitude of the stable flame front, using the assumption
that the flame front is composed of joined parabolic segments whose
amplitude is small when compared to their wavelength \cite{Zeldovich85}.
The resulting estimate depends on the growth rate, 
Eq.~(\ref{eq_growth_thin_front}),
\[
  \frac{h_T}{l} = \frac{1}{8}\;\left( \gamma\frac{l}{s_\circ} \right)\, ,
\]
Comparing the result with the fit derived from the experimental data
shown in Fig.~\ref{fig_ht} we notice that, in the thin front limit
and for values of $G$ larger than critical, both numerical experiment
and theoretical model predict $h_T / l \approx 0.22 \sqrt{LG}$.

Finally, we note that a quick comparison of the asymptotic behavior of
the Rayleigh-Taylor \cite{Chandrasekhar61,Landau87} and
Landau-Darrieus \cite{Darries38,Landau44} instabilities for
large $L$ gives $\gamma \propto L^{-1/2}$ for Rayleigh-Taylor and
$\gamma \propto L^{-1}$ for Landau-Darrieus. In our Boussinesq case,
the same asymptotic limit gives $\gamma \propto L^{-1/2}$: the
instability behaves like the Rayleigh-Taylor instability at long
wavelengths, i.e., longer wavelengths grow more slowly, but saturate
later and reach larger front speeds.


\subsection{Transition to the travelling wave}

\begin{figure}[b]
\begin{center}
\epsfxsize=\plotwidth
\mbox{\epsfbox{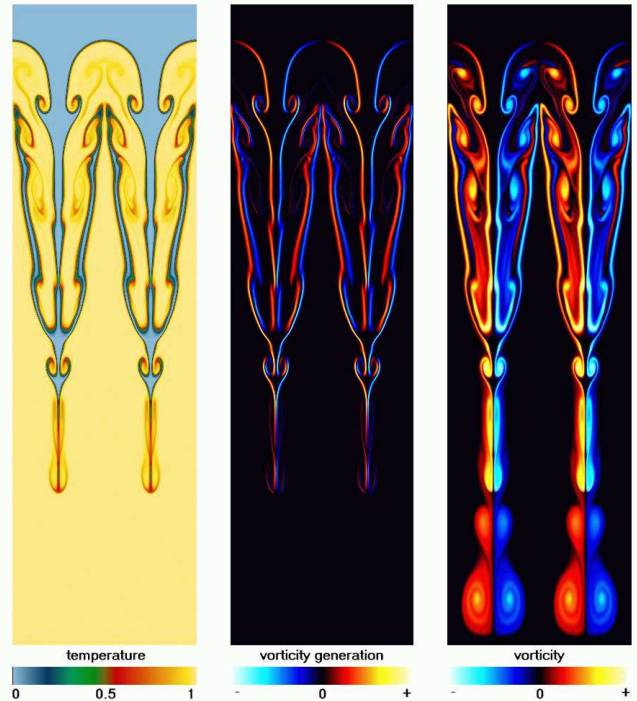}}
\end{center}
\caption{
  The temperature, vorticity generation rate, and vorticity (from left
  to right) for the system with $L=256$ and $G=4$ at time
  $t=72\,\delta/s_\circ$. The initial amplitude of the perturbation was
  $a/l=1/8$. }
\label{fig_swirly}
\end{figure}

The transition time during which the temperature front is formed is of
the order of the laminar burning time across the period,
$0.5\,l/s_\circ$, but a much longer time is needed to stabilize the
velocity pattern behind this front.
Fig.~\ref{fig_interface_similarity} illustrates the process for a
moderate value of $L$; in Fig.~\ref{fig_swirly} we show snapshots for
a flame with an $L$ value closer to the Rayleigh-Taylor limit just
discussed. Indeed, Fig.~\ref{fig_swirly} shows morphology strongly
reminiscent of the Rayleigh-Taylor instability, namely upward-moving
``bubbles'' and downward moving ``spikes''. As mentioned earlier, the
reaction stabilizes the shape of the moving front, and eventually the
flame interface will become smooth, similar to those shown in
Fig.~\ref{fig_levels}. The typical flame stabilization time is of the
order of $L/s_\circ$, provided the initial perturbation amplitude is
large enough (on the order of a fraction of $L$). During the
transition, the system with larger $L/\delta$ ratio develops more
complicated structures (compare Fig.~\ref{fig_swirly} with
Fig.~\ref{fig_interface_similarity}) --- the details on the scale of
flame thickness and smaller are consumed by the burning. A similar
effect is observed in the Rayleigh-Taylor instability on the
dissipation scale, but due to viscosity and diffusion rather than
burning.

\begin{figure*}[t]
\begin{center}
\epsfxsize=2\plotwidth
\mbox{\epsfbox{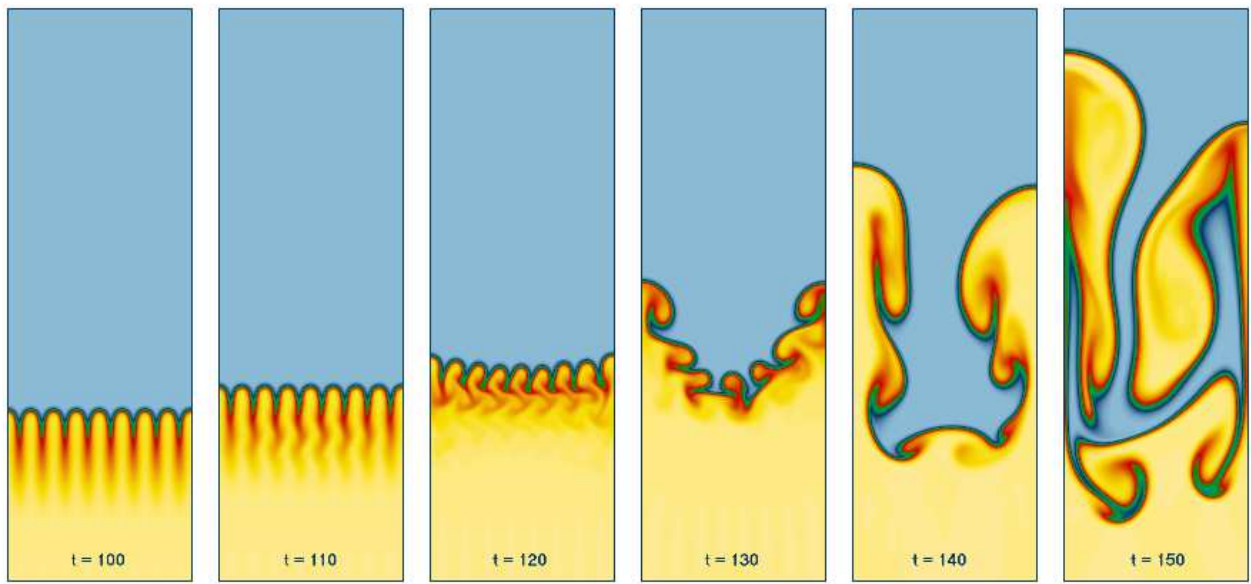}}
\end{center}
\caption{
  Symmetry breaking due to numerical noise and resulting instability.
  The snapshots are taken at times
  $ts_\circ/\delta=100,110,120,130,140,150.$ }
\label{fig_multi}
\end{figure*}

The images shown in Fig.~\ref{fig_multi} illustrate the propagation of
a flame with eight wavelengths (with $L=16$, $G=4$) within the
computational box with reflecting boundary conditions. The chosen
parameters place the system well inside the unstable regime, and, by
the time $t \approx 30\,\delta /s_\circ$ the system forms the curved
travelling wave solution with wavelength $L=16$. This solution is
exactly the same as the curved solution obtained in the
half-wavelength computational box, propagates with the same speed, and
remains unchanged until time $t \approx 100\,\delta/s_\circ$. (We note
that the ``wall effect" seen in this figure reflects both the presence of the 
walls (and choice of boundary condition at the walls) and the choice of
phase for the initial perturbation.)

The symmetry of the initial conditions requires zero horizontal
velocity at $x=nl/2$, $n=0,1,2,...$; this symmetry constraint is
clearly broken for $t\gtrsim 100 \delta/s_\circ$, and the travelling
wave solution becomes violently unstable. The cause of this symmetry
breaking is apparently accumulated numerical errors (noise) in the
calculation. The source of this noise is the iterative solution for
the stream function, so the noise has the wavelength of the
computational domain. Since perturbations with larger wavelengths move
faster, the system will eventually pick the travelling wave
configuration corresponding to the largest possible wavelength --- in
the example shown, the wavelength $L=256$ (twice the box size).

The instability shown in Fig.~\ref{fig_multi} is not related to
metastable behavior near $\Gcr$ discussed in \cite{Berestycki02} ---
both wavelengths present in the system are unstable for $G=4$. Rather,
this simulation is an illustration of the fact that the small
wavelengths have faster initial growth rates, but saturate at lower
speeds. As a result, the instability exhibits a strong inverse
cascade. More careful modelling of the noise introduced to the system, 
as well as more realistic treatment of boundary conditions at the walls,
will be necessary to learn about the instability dynamics in an infinitely
large domain; in particular, we believe that periodic boundary conditions 
should be imposed at the walls in order to study this problem further.
In such a system we would expect unbounded growth of
instability size; in a natural system we would expect the upper bound
to be set by extrinsic spatial scales of the physical system.


\section{Summary and Discussion}

In this paper, we have studied the fully nonlinear behavior of
diffusive pre-mixed flames in a gravitationally stratified medium,
subject to the Boussinesq approximation. Our aim was both to compare
our results for a viscous system with analytical (and empirical)
results in the extant literature, and to better understand the
phenomenology of fully nonlinear flames subject to gravity.

The essence of our results is that the numerics by and large confirm
the Markstein and Pelc\'e \& Clavin models, and extend their results to
finite viscosity. We have shown explicitly that there is an extended
regime for flames with finite flame front thickness for which the
scaling on the $LG$ product applies (as it is known to do in the thin
flame front limit). We have also examined the details of the flame
front structure, and are able to give physically-motivated
explanations for the observed scalings, for example, of the flow
length scale behind the flame front on Prandtl number.

We have also observed a potentially new instability, which arises when
noise breaks the symmetry constraint of the initial front
perturbation. Our study suggest that this instability differs
significantly from finger merging behavior of the non-linear
Rayleigh-Taylor instability, in which the finger merging process
resembles a continuous period-doubling phenomenon (e.g. adjacent
fingers at any given generation merge in pairs). In contrast, the
instability we observe seems to involve seeding, and strong growth, of
modes with wavelengths much larger than the wavelength of the dominant
front disturbance. We are currently investigating this instability in
greater detail.

Finally, it is of some interest to consider the implication of our
results for astrophysical nuclear flames, as arise in the context of
white dwarf explosion. Using the results of Timmes \& Woosly
\cite{Timmes92}, we find that we would be far into the thin flame
limit, with a density jump at the flame front $\Delta \rho \sim 0.1
\rho$; hence our Boussinesq results are rather marginal in their
applicability. Nevertheless, one can ask what the expected flame speed
up would be in this limit; using our results we find that $s/s_\circ
\approx (1+0.0486 \, LG)^{1/2}$, with
$LG=(\Delta\rho/\rho)\,lg/s_\circ^2$. Using the length scale of the
order of a fraction of white the dwarf radius, $l \approx 10^3 km$,
gravitational acceleration on the surface of the star, $g \approx 10^3
km/s^2$, and the laminar flame speed given by \cite{Timmes92},
$s_\circ \approx 100 \, km/s$, we obtain $LG \approx 10$, and
consequently a speed up of $s/s_\circ\approx 1.2$. Smaller laminar flame
speeds would lead to the flame velocities independent of the laminar
flame speed, $s=0.23 \, (lg\Delta\rho/\rho)^{1/2}\approx 100 \, km/s$,
which could be also derived using the rising bubble model
\cite{Layzer55}.  Evidently, the flame speedup in this limit is very
modest. Whether compressibility has much effect on this conclusion
remains to be established and is now under active investigation.

\section{Acknowledgements}

This work was supported by the Department of Energy under Grant
No.B341495 to the Center for Astrophysical Thermonuclear Flashes at
the University of Chicago.


\end{document}